\documentclass[preprint2]{aastex}

\def\etal{\mbox{et al.\/}}

\def\m{$^{\rm m}$}

\def\lae{\mathrel{<\kern-1.0em\lower0.9ex\hbox{$\sim$}}}
\def\gae{\mathrel{>\kern-1.0em\lower0.9ex\hbox{$\sim$}}}
\def\m{Mrk~421\ }

\def\euve{\mbox{\em EUVE\/}}

\def\pap1{\mbox{paper~{\sc i}\/}}

\slugcomment{Submitted to {\it The Astrophysical Journal}}

\shorttitle{Cagnoni, Papadakis \& Fruscione}
\shortauthors{Extreme Ultraviolet Lightcurves of \m}

\begin{document}

\title{Four Years of Extreme Ultraviolet Observations of Markarian 421.
II: Temporal Analysis}
\author{I. Cagnoni\altaffilmark{1,2}, I. E. Papadakis\altaffilmark{3} \and  A. Fruscione\altaffilmark{2}}
\affil{$^{1}$ SISSA, Via Beirut 4 -34138, Trieste, Italy}
\affil{$^{2}$ Harvard-Smithsonian Center for Astrophysics, 60 Garden Street, 
Cambridge, MA 02138, USA}
\affil{$^{3}$ Physics Department, University of Crete, 710 03
Heraklion, Crete, Greece }
\email{ilale@sissa.it}

\begin{abstract}

The {\it Extreme Ultraviolet Explorer} ({\it EUVE}) 
satellite accumulated $\sim$ one million seconds of public data 
between 1994 and 1997 for the BL Lacertae object Markarian 
421. This is the second of two papers in which we present the results of 
 spectral and temporal analysis of this {\it EUVE} data set.
We analyze in the present paper the imaging data by means of power spectrum and structure function techniques, while the spectral analysis is presented 
in a companion paper.
We find for \m \/ a power spectrum with slope $-2.14\pm 0.28$ with a break
at $\sim 3$ days. This is the first time that a break in the power spectrum 
of a BL Lacertae object has been found.
We also find  evidence of non-stationarity for \m \/ EUV emission.

\end{abstract}


\section{Introduction}
BL Lacertae objects are an extreme subclass of Active Galactic Nuclei
(AGNs) emitting highly variable non-thermal radiation over 20 decades
of frequency from radio to
TeV energies.  The mechanism responsible for the production of
radiation over such a wide range is thought to be synchrotron emission
followed by inverse Compton scattering from energetic electrons.  Some
of the extreme properties of BL Lac objects, such as rapid
variability, high gamma-ray luminosities and superluminal motion (e.g. Urry
\& Padovani 1995), are explained invoking relativistic beaming of the
emitted radiation along the jet axis which is taken to be oriented at 
small angles to the line of sight.

\m is one of the closest ($z$=0.0308) and best studied BL Lacertae
objects. It is one of the strongest TeV sources and one of the
brightest extragalactic EUV sources.  The EUV emission of \m  
during quiescence is believed to be the peak of a synchrotron
spectrum. 
\m was monitored by the {\it EUVE} satellite for four years
(1994,1995,1996 and 1998) as part of multiwavelength campaigns and it
was serendipitously observed in 1995 and in 1997 with the {\it EUVE}
photometers for a total of one million seconds, the best 
sampled light curve for any BL Lac object at high energies\footnote
{As in paper~{\sc i}, we are not considering the most recent
observation (April 1998) since the data are presently being analyzed by the
original proposers (Marshall \etal\ in preparation)}. 
Table~1 summarizes all the {\it EUVE} observations.

The main purpose of the present paper is to analyze in an homogeneous way the
1100~ks of {\it EUVE} public imaging data for \m (less than half of
it has been analyzed to date, Fruscione et al. 1996, Kartje et
al. 1997) in order to perform the first detailed variability
analysis (both in terms of power spectrum and structure function
analysis) for a BL Lacertae object in the EUV.  In general,
variability studies by means of power spectrum analysis can reveal
characteristic time scales of the emission mechanism
(e.g. periodicities) and can constrain physical models.  
Our analysis reveals the existence of
a characteristic timescale for \m that can constrain current models.\\ 
\indent The
paper is organized as follows: all the {\it EUVE} lightcurves of \m
collected from 1994 to 1997 are presented in Section 2.  In Sections 3
and 4 we present the results from the power spectrum and structure
function analysis, respectively. In Section 5 we discuss those results.
Finally in Section 6 we give some conclusionary remarks. For the energy
spectrum data and analysis we refer the reader
 to the companion paper Cagnoni \&
Fruscione (2000, hereafter paper~{\sc i}).

\section{Observations}

A general description of the {\it Extreme Ultraviolet Explorer}
satellite (\euve) is given in paper~{\sc i}.  The photometric
instruments on board {\it EUVE} are the Deep Survey/Spectrometer,
(DS/S, covering the 68-178 \AA\ band for photometry; Welsh et al. 1990) 
used for the deep EUV survey along the ecliptic
and for pointed spectroscopic and imaging observations, and three
co-aligned photometers (the ``scanners''), observing in four different
bandpasses (from 58 to 740 \AA) and mounted orthogonally to the DS/S. 
The photometers were used to
carry out the \euve\ all-sky survey and are now employed in the
``Right Angle Program'' (i.e. the observation of serendipitous targets
while the DS/S is conducting the primary science
observation). Figure~\ref{fig1} shows a comparison of the DS/S and scanner
effective areas.

\m was observed several times from 1994 to 1997:
four times with the DS/S and twice with the scanners; Table~1 
summarizes these observations.
Because of interstellar medium absorption along the line of sight
($N_H=1.45 \times 10^{20}$~cm$^{-2}$, Elvis, Wilkes
\& Lockman 1989), \m was detected only in the Lexan/B filter (the shortest
wavelength bandpass) both in the
scanners (58-174~\AA) and in the DS/S (67-178~\AA).


\subsection{Serendipitous sources}

Two faint serendipitous sources are clearly visible next to \m\ in the DS
image. We identified the closest one
($\sim 2^{\prime}$ from \m) with 2 unresolved stars in
a double system: HD~95934 ($M_{V}=6.00$, Spectral type A3III-IV) and
BD+39~2414B($M_{V}=12.6$). The other one at $\sim 4^{\prime}$ from
\m was identified with another star in double system: HD~95976
($M_{V}=7.4$, Spectral type FII).

The DS 50\% energy radius is about $0.3^{\prime}$ up to
$0.3^{\circ}$ off-axis. In scanner A, this half energy radius is 
$\sim3^{\prime}$ at $\sim 1.5-2^{\circ}$ off-axis (where the sources
lied in the 1995 and 1997 observations) and the sources are not
resolved.
During the analysis of DS data we excluded a circle centered on
HD~95934; this excluded only a negligible fraction of counts from \m.
%

\subsection{Light curves}

We extracted the light curves from the DS or Scanner
time-ordered event lists using the {\it EUVE} Guest Observer Center
software (IRAF/EUV package) and other IRAF timing tasks adapted for
{\it EUVE} data.  All the lightcurves were binned over one average {\it
EUVE} orbit (5544~s), but effective exposure times per bin vary due
to several correction factors different case by case 
(e.g. instrumental deadtime, telemetry saturation, high particle
background, dead spot).

Unless described otherwise, we chose the
extraction region for the source as a circle with a 2.5$^{\prime}$
radius excluding a circle with a 0.67$^{\prime}$ radius centered on the
serendipitous source.
The background region is an annulus with inner and
outer radii equal to $\sim 4^{\prime}$ and $8^{\prime}$ respectively
which avoids the other serendipitous source.

In order to compare scanner and DS data, we had to normalize the count
rates.  Count rates depend on the effective area of the telescope and
on the spectral shape of the source.  We convolved a sample of input
spectra (absorbed power law with slopes as in Table 2 of \pap1) with
the DS and scanner effective areas and defined the normalization
factor as the ratio of peak of the derived count rates (see Figure~\ref{fig2} as
an example for $\alpha=2$). The
average normalization factor is about 1.4$\pm$0.3 (depending on the
input spectrum). We used this value throughout the paper and verified
that the results in the power spectrum analysis are all consistent
within the errors of this value.

The combined light curve of all the observations is presented in Figure~3.
In this section we will give a qualitative description of each
lightcurve in terms of flares and sub-flares: a general
rising/decreasing trend will be addressed as a ``flare'', while faster
variations superimposed on the general trend will be referred  to as
sub-flares.  This description is useful in understanding the results from the power
spectrum and structure function analysis.

\subsubsection{1994 DS light curve}


Figure~\ref{fig4} shows the 1994 light curve, which was obtained eliminating
time intervals when the source crossed the dead spot
\footnote{A small region of reduced
gain and detector quantum efficiency near the center of the DS
detector, caused by the observation of the very bright EUV source
HZ~43 (Sirk 1994)} 
(about 1440 s).

The source dimmed
for about 3 days, brightened again for another $\sim 3.5$
days, until it reached a level of $\sim 0.2$ counts/sec
and stayed there for the remaining of the observation (another
$\sim 3.5$ days). 
This light curve behavior could be sampling the end of a flare
and the beginning of a new one.  Interestingly, when the source
increases its flux, after MJD~49447, (i.e. possibly during the beginning of a
new flare) smaller ``sub-flares" are superimposed on it; this trend
is also present in the 1995 scanner lightcurve 
(see 2.2.2 and Figure~\ref{fig5}).

\subsubsection{1995 scanner A light curve}

During this scanner exposure optimized for the primary scientific DS/S
observation of HD~41511, \m partially falls in the region usually
discarded by the standard analysis pipeline because of the obscuration
due to the filter support and because of the high particle background
close to the detector edge.  We extracted the lightcurve only from the
``clean'' region of the detector and we
corrected the count rate for the fraction of the count lost
($\sim$10\%).  The background region was chosen in the same detector
quadrant as the source and large enough to match the ratio
background/source ($\sim 7.8$) region used for the other lightcurves.
The lightcurve
is plotted in Figure~\ref{fig5}; the open circles represent the count rate
measured in scanner A, while the filled circles are the count rate
normalized to the DS; this lightcurve has as behavior similar to the
1994 DS lightcurve: when the source brightens, after MJD~49754,
``sub-flares'' appear above the main trend.

\subsubsection{1995 DS light curve}

During this DS/S observation, the source was pointed 0.3$^{\circ}$ off axis
both to avoid the dead spot and to extend the spectrum toward shorter
wavelengths (see \pap1).
The light curve is shown in Figure~\ref{fig6}.  

The behavior of the source during the DS 1995 observation is complex:
at the beginning \m brightens by  a factor of $\sim 65$\% in less than 2 days and
this flare, which lasts for about 3 days, is well correlated to a
TeV/X-Ray flare (Buckley \etal\ 1996).  Right after the first one,
there is a second flare lasting for $\sim 6$ days which can also be
seen as the superposition of two separate flares (the second one
starting at $\sim$ MJD 49838). After MJD 49845.5, the source rises
from $\sim0.20$ counts s$^{-1}$, up to $\sim 0.28$ counts s$^{-1}$, 
and then decays to $\sim 0.23$ counts s$^{-1}$,
within $\sim 3$ days.  The overall variability of the source is about
a factor 3.

\subsubsection{1996 DS light curves}

In 1996 \m was observed few days before and few days after the detection of
a large TeV flare on May 7th 1996 (Zweerink \etal\ 1997).
The lightcurves are shown in Figure~\ref{fig7}.
This lightcurve looks different from all the other ones 
because no slow variations seem to be
present; the source appears to be at a constant level 
with lots of small amplitude, fast variations
superimposed on it; i.e., the light curve appears to be the sum of
flares which live a short time, and rise or decay faster when compared
with the flares seen in the other light curves (see for example the
two flare like events after MJD=50200).

\subsubsection{1997 scanner A light curve}

In 1997 \m was observed for 4 days with the scanner A telescope during
the DS pointed observation of $\Gamma$~Tauri; the source was detected
within the useful area of the scanner A field of view allowing an
easier data reduction than the previous scanner observation.  The
light curve is shown in Figure~\ref{fig8}: the open circles represent the
scanner data while the filled circles correspond to the count rate
normalized to the DS.  This is the shortest lightcurve and its shape
suggests that we may be looking at the tail of a flare and at the
beginning of a new one.

\section{Power Spectrum Analysis}

In order to quantify the variability seen in the \euve\ lightcurves of
\m, we computed their periodogram as:

\begin{equation}
\hat{I}(\nu_{i})=\frac{\Delta t}{N}\{ \sum_{i=1}^{N}  
[x(t_{i})-\bar{x}]e^{-i2\pi \nu_{i} t_{i}} \}^{2}-C,
\end{equation}

where: $\bar{x}$, $\Delta t$ and $N$ are the mean value, bin size 
and number points of each light curve
respectively, and $\nu_{i}=i/(N\Delta t)$, $i=1,2, ..., (N/2)-1$ 
(e.g. Deeming 1975). 
The constant $C$ in
equation (1) represents the power level due to Poisson statistics,
$C=\Delta t \sigma^{2}_{noise}=(\Delta t/N)\sum_{i=1}^{N}err(i)^{2}$
($err(i)$ is the error of each light curve point).
The subtraction of this
constant level is necessary since we are interested in the source power
spectrum only and the {\it EUVE} light curves were obtained
with two different instruments, hence the Poisson statistics will be
different in them.
The periodogram calculated in this way has the units of (counts/sec)$^{2}$/Hz,
and its integral over positive frequencies is equal to half the light
curve variance.

It is common to average the periodogram estimates in order to reduce the
variance and approximate their probability distribution function to a
Gaussian. Furthermore, it is better to estimate the logarithm of the power
spectrum instead of the power spectrum itself, since the logarithmic
transformation brings the distribution function of the power spectrum
estimates closer to a Gaussian (Papadakis \& Lawrence, 1993). Therefore,
we calculated the logarithm of the binned periodogram estimates,

\begin{equation}
\log_{10} [ \hat{P}(\nu_{i})] =\log_{10}[ \frac{\sum_{j}
\hat{I}(\nu_{j})}{m}],
\end{equation}

(where $\nu_{i}=\sum_{j} \nu_{j}/m$, and $\nu_{j}=j/(N\Delta t)$) and
used it as our estimate of the {\it logarithm} of the power spectrum. The
index $j$ varies over $m$ consecutive periodogram estimates and the index
$i$ varies over the $[(N/2)-1]/m$ subsets into which we have grouped the
periodogram estimates. Taking into account the propagation of errors (e.g.
Bevington 1969), the error on the logarithm of the power spectrum is

\begin{equation}
\sigma(\log_{10} [ \hat{P}(\nu_{i})])=
\log_{10}(e)\sigma_{\hat{P}(\nu_{i})}/\hat{P}(\nu_{i}),
\end{equation}

where $\sigma_{\hat{P}(\nu_{i})}$ is the square root of the variance of
the $\hat{I}(\nu_{j})$' s around the mean value, $\hat{P}(\nu_{i})$, at
each bin. 

\subsection{Power spectra of the orbit-binned lightcurves}

We estimated the power spectrum of the DS and scanner A light
curves binned over one average {\it EUVE} orbit (5544 s) shown in
Figures~\ref{fig3} -- \ref{fig8}. 
To avoid the gap in 1995 DS observation (Figure~\ref{fig6}) we
divided it into 2 parts and performed a power spectrum analysis on each
one of them. Using equations~1 and 2, we estimated the logarithm of the
power spectrum with $m=10$. A larger value for $m$ would bring the
distribution function of the power spectrum estimates closer to a
Gaussian (Papadakis \& Lawrence 1993). 
However this small value of $m$ was necessary in order to have a
reasonable frequency resolution, i.e. in order to have the maximum possible
number of power spectral points above the Poisson noise level.

The power spectra of all the light curves, normalized to the square
mean value of the respective light curve, are shown in Figure~\ref{fig9}.  
The normalisation is
necessary if one wants to compare the power spectrum amplitudes, since the
light curves were obtained from two different instruments.
All the power spectra show a similar red noise behavior 
at frequencies $< 10^{-4.3}$Hz (i.e.  all of
them rise logarithmically towards lower frequencies in a similar way),
 except for DS 1996 observation
which seems  to flatten for $\nu <10^{-5}$~Hz.

We fitted the
DS power spectra with a simple power law model of the form $P(\nu) = A
\times (10^{-5}\nu) ^{-\alpha}$ (the scanner A power spectra had too
few points to fit a model to them). In this model, the normalization
$A$ is equal to the value of $P(\nu)$ at $\nu=10^{-5}$ Hz.  The model
fit was done using the power spectra before normalizing to the mean
square value. The best model fit results are listed in Table~2. The
model gives a good fit to all the DS power spectra, but, due to the
small number of points, the errors on the model parameters are large.
The best fit values of $A$ are within the errors for all the DS
spectra. However, the best fit slope values show a larger scatter. In
particular, the difference between the DS 1995 (part 1) and the DS
1996 best fit slope values is $2.7\pm 1.0$. This $\sim 3\sigma$
difference indicates that the shape of the power spectrum does not
remain constant with time, i.e.  that the EUV emission of \m is not
stationary \footnote{A lightcurve is said to be stationary if its
mean, variance and power spectrum remain constant in time
(e.g. Chatfield 1989); see Section~5.4 for further details}.  However,
looking at DS 1996 power spectrum (Figure~\ref{fig9}), there is a
possibility that the flatter slope obtained from the fit is due to a
flattening around $\nu=10^{-5}$ Hz.  If this is the case, a broken
power law model should give a better fit to the data. However, with
just 5 points, it is not possible to investigate if a broken power
improves the fit to the data in a statistically significant way (e.g. by
the use of an F-test). On the other hand, perhaps the flattening of
the power spectrum is a spurious effect introduced by the small
binning used ($m=10$). In an attempt of ruling out this possibility we
used a larger value ($m=20$).  Figure\ref{fig10} compares the results:
open circles for $m=10$ (as in Figure~\ref{fig9}) and filled circles
for $m=20$. Both power spectra look similar, and the lack of frequency
resolution in the latter case cannot allow us to reach a firm conclusion.

Despite this possible non-stationarity, it is useful to combine the
periodograms of all the DS and scanner A light curves in order to
increase the number of power spectrum estimates (and hence the
frequency resolution).  We first normalized the data sets to the
square mean value of the respective light curve and then combined them
into one dataset by sorting them in order of increasing
frequency. Then we grouped this ``combined" periodogram into bins of
size $m=20$ since the larger number of estimates allows us to use a
larger bin size value.  Figure~\ref{fig11} shows a plot of the
resulting power spectrum (filled circles). Note that the combined
power spectrum has the units of Hz$^{-1}$.  This spectrum can be
considered as the ``average'' EUVE power spectrum of the source. A
power law model gives a good fit to this average power spectrum (solid
line in Figure~\ref{fig11}; Table~2) with a best fit slope value of
$2.14\pm 0.28$.

\subsection{Power spectra of the 500~s binned lightcurves}

Due to the indication of a possible break in the power spectrum and/or
a possible non-stationarity in the {\it EUVE} light curve of MRK~421,
we decided to investigate further the power spectra of the DS 1995 and
1996 light curves.
Our main aim was to determine the high frequency shape of the power
spectrum as well as possible.  For this reason, we used a $\sim10$
times finer binning, with a bin size of 500~s, for the light curves;
we were thus able to extend the power spectrum to higher frequencies
and to use a larger value for $m$ ($m=20$ in equation 2) to bring the
distribution function of the power spectrum estimates closer to a
Gaussian.

Figure 12 shows the DS 1995 and 1996 power spectra estimated
with equation (1), without subtracting the expected Poisson noise
level, and binned as in equation (2) with $m=20$; the long dashed
lines in these plots indicate the Poisson noise power level $C$.  Both
power spectra show a large peak at $\sim 1.8\times 10^{-4}$ Hz
corresponding to a period of 5556 s, i.e. the \euve\ orbital
period. The next strongest peak is at a harmonic of that frequency.
These peaks are due to the uneven sampling of the 500~s binned light
curves.




Apart from those peaks, the two power spectra appear to have a similar shape: 
they rise towards low frequencies, and they are flat at higher
frequencies. So, it seems as if the use of a smaller bin size in the light
curves has not helped us much. The power spectra, at most frequencies,
appear to be flat and dominated by the Poisson noise power. However, while
the DS 1995 estimates at high frequencies are close to the expected
Poisson power level, the DS 1996 power spectrum is higher than this level. 

We fitted both spectra with a modified power law model with one of the
slopes fixed at 2.14 (see below).  Due to the effects of the window
function, it is not possible to fit the logarithm of the model
directly to the power spectrum calculated from the 500 sec binned
light curves.  Instead, we convolved the model, let us say $M(\nu)$,
with the window function at each frequency $\nu_{i}$,

\begin{equation}
 I_{mdl}(\nu_{i})=\frac{\Delta t}{N}\int_{-\infty}^{+\infty}
M(\nu)W(\nu_{i}-\nu)d\nu.
\end{equation}

The function $I_{mdl}(\nu_{i})$ takes account of the window function
effects. We call $I_{mdl}(\nu_{i})$ the ``convolved model''.

As $M(\nu)$, we used the following function,

\begin{equation}
M(\nu)=\frac{A}{[(2\pi\nu)^{2}+f_{b}^{2}]^{\alpha}}+C,
\end{equation}

which has a power law form with a slope of $2\alpha$ at frequencies
$>>f_{b}/2\pi$, and flattens to zero slope at frequencies $<< f_{b}/2\pi$
(as before, $C$ represents the power level due to the Poisson counting
statistics). Note that as $M(\nu)$ we did not use a simple power law model
because in this case it is not easy to estimate the convolution model
(equation~4) analytically (a simple power law model cannot be defined at
$\nu=0$) and because there might indeed be a break, as suggested by DS 1996 
power spectrum (Figure~\ref{fig9}).

Then, we averaged the convolved model $I_{mdl}(\nu_{p})$ over groups of
size $m=20$ exactly as we did when estimating the power spectrum.
The logarithm of the convolved model power spectrum becomes:
 
\begin{equation}
\log_{10}[P_{mdl}(\nu_{i})]=\log_{10} [\frac{\sum_{j}
I_{mdl}(\nu_{j})}{m}].
\end{equation}

We can now fit $\log_{10}[P_{mdl}(\nu_{i})]$ to the estimated power
spectrum, $\log_{10}[\hat{P}(\nu_{i})]$. In this way, we are taking into
account both the window function and any effect that the binning procedure
may have on the power spectrum estimation. 

For the fit we used standard $\chi^{2}$ statistics, we kept $2\alpha$ 
fixed to the value 2.14, and  we did not allow $C$ to vary but
rather used the value found as described at the beginning of Section
3.  Table~3 summarizes the best fit results for each observation; the
errors quoted represent the 90\% confidence levels assuming that $A$
and $f_{b}$ are correlated. By far the largest $\chi^{2}$ value is
for the DS 1996 model fit. However, it is not straightforward to judge
the goodness of fit using the these $\chi^{2}$ values. Due to the
uneven sampling of the light curves, the periodogram estimates are no
longer independent. Therefore, the errors on
$\log_{10}[\hat{P}(\nu_{i})]$ may be larger than those estimated (and
hence the $\chi^{2}$ values smaller). At the same time, depending on
the amount of correlation between the $\hat{I}(\nu_{i})$'s, the
$\log_{10}[\hat{P}(\nu_{i})]$ estimates may be correlated as well. As
a result, the number of degrees of freedom may be smaller than those
listed in Table~3.

The best model fits are also plotted in Figure 12 with a solid line; 
these plots show that the model agrees well with the DS 1995 power
spectrum, but not with the DS 1996 (as expected from the large $\chi^{2}$
value). We fitted again the DS 1996 power spectrum with the model defined
in equation~(5) but we let the slope as a free parameter this time. The
best fit results are listed in Table~3 (values in a parenthesis), and the
best fit model is also plotted in Figure~12 (b) (dotted line). As this figure
shows, the new model fit agrees better with the power spectrum. The
best fit slope is now $1.1 \pm 0.1$ significantly smaller than the average
slope of $2.14 \pm 0.28$. The $\chi^{2}$ value is also reduced by 74.9. 
The ratio of the reduced $\chi^{2}$ values for the two fits is 2.1 (note
that this ratio does not depend on whether the errors on the power
spectrum are underestimated or not). Using the number of degrees of
freedom listed in Table~3 and the F-test, we find that the improvement on
the fit when we let the slope to be a free variable is highly significant
(the probability of a significantly better fit when $\alpha=1.1$ is larger
than $99.5\%$). The probability remains larger than $95\%$, even if the
degree of freedoms listed in Table~3 are reduced by a factor of two.

Furthermore, to check for any other possible indication of
non-stationarity, we performed a fit to the power spectra of all the
500 sec binned light curves (including the scanner A light curves)
using the model defined in equation~(5) with $2a=2.14$ . The best
fit model results are listed in Table~3, and the best fit model,
together with the power spectra, are plotted in Figure~13 (the dotted
lines in this figure show the Poisson noise power level). This figure
shows that the model gives a good fit to all the power spectra.

We conclude that all the {\it EUVE} power spectra of \m are consistent
with a power law model with a slope of $\sim 2.1$. The only
exception is the power spectrum of the DS 1996, which shows a
significantly flatter slope of $1.1\pm 0.1$. The clear change in slope
in the 1996 data shows that
the EUV emission of \m is not statistically stationary (see Section~5.4).

As hinted by Figure~\ref{fig9} for DS 1996 (open circles), 
from the fit of the $500$~s binned power
spectra we found indication for the presence of a low frequency turn
over in the {\it EUVE} power spectrum of \m below which the power
spectrum flattens. This finding, if confirmed, would be the first
detection of a break in the power spectrum of a BL Lac object.  Since
we cannot estimate the power spectrum at frequencies lower than $\sim
10^{-5}$ Hz, using the value $m=20$,
we cannot be certain of the slope below the turn over
frequency. All we can say with the present data is that the power
spectrum should show a change in its slope at $f_{b}$. Furthermore the value of
this knee frequency is not constrained very well from our data.  Using the
best fit results from the DS spectra only (as they have the largest
number of data points), we get a straight mean value of $f_{b}=4\times
10^{-6}$ Hz. By simply combining the overlapping $90\%$ confidence
limits on the best values of $f_{b}$, we find that with probability
larger than $68\%$, $f_{b}$ should be between $7\times 10^{-7}$ Hz and
$8\times 10^{-6}$ Hz.


From a physical point of view, it is natural to expect that low
frequency knees should be present in the power spectra of AGN. Lack of
this turnover would imply a physical system with infinite variance, a
system whose properties change continuously with time.  The frequency
turnover is not evident in the combined {\it EUVE} power spectrum of
\m (Figure~\ref{fig11}), because that spectrum does not extend to frequencies
below $\sim 10^{-5.5}$ Hz. Its presence is implied only when we take
proper account of the window function effects in the model fitting of
the 500~s light curve power spectra. If the frequency turn over were
located at a much lower frequency, more power would be transfered to
higher frequencies.  The resulting power spectrum would have a higher
normalization than the observed one. 
To put in other words, the amplitude of the variations in the light
curves would be larger, i.e. they should have a higher variance.  For
example let us assume that the source power spectrum is well describe by
equation (5) and that the break is at $10^{-7}$ Hz. Using equations (4)
and (6) we can predict how the power spectrum of the DS 1995 light curve
should be in this case (dotted line in Figure~12a). As this Figure shows,
the power spectrum should have a much larger amplitude than the one that
is actually observed.

Due to the significance of such a feature in the power spectrum of the
source, we searched further evidence of its existence by means of a
structure analysis of the light curves. We present below the results
from this analysis of the {\it EUVE} light curves which confirm the
existence of this turn over in the power spectrum of \m.

\section{Structure Function Analysis}

The Structure Function (SF hereafter) is defined as,

\begin{equation}
SF(\tau)=2\sigma^{2}-R(\tau),
\end{equation}

where $\sigma$ and $R(\tau)$ are the variance and auto-covariance at lag
$\tau$ of the light curve (equation~A4 in  Simonetti, Cordes \& Heeschen, 
1985). For a given light curve, it is usually estimated by,

\begin{equation}
\hat{SF}(\tau)=\frac{1}{N}\sum_{i=1}^{N-\tau}[x(i+\tau)-x(i)]^{2},
\end{equation}

(equation~A8 in Simonetti, Cordes \& Heeschen, 1985).  The SF
gives a measure of the mean difference in the flux of two light curve
points, as a function of their time separation (lag) $\tau$. There is a
simple correspondence between the SF and the power spectrum: when
$P(\nu)\propto \nu^{-\alpha}$ then $SF(\tau) \propto \tau^{\alpha-1}$. 
Furthermore, if the power spectrum flattens below a frequency $f_{b}$,
then $SF(\tau)\sim 2\sigma^{2}$ in the limit $\tau >> 1/f_{b}$.

The main advantage of the structure function analysis is that the SF
estimation is not affected by gaps of missing points in the light
curve (it is easy to show using equation~(8), that $<\hat{SF}(\tau)> =
SF(\tau)$, irrespective of the light curve sampling pattern). However,
it is difficult to estimate the errors on $\hat{SF}(\tau)$ because, as
with estimates of other time domain quantities (eg auto,
cross-correlations), the SF estimates at different lags are heavily
correlated (much more so than the periodogram points at different
frequencies, even in the case of unevenly sampled data).

To estimate the SF of MRK~421, we used together all the orbit-binned
DS and scanner A light curves and constructed a single, long {\it
EUVE} light curve of the source (in order to use the scanner data into
an overall light curve, we normalized them to the DS expected count
rate as explained in Section~2.2). For each pair of data points in the
light curve, $x(t_{i})$ and $x(t_{j})$, we computed their time
difference, $\Delta t_{i,j}=t_{i}-t_{j}$, and their flux difference,
$FD_{i,j}=x(t_{i})-x(t_{j})$. Let us assume that we have chosen a lag
bin of size $\Delta \tau$, and that there are $n$ points with $\Delta
t_{i,j}$ in the interval $[\tau-\Delta\tau/2, \tau+\Delta\tau/2]$. We
used the average value of the $n$ $FD_{ij}$ measurements, $\sum
FD_{i,j}/n$, as an estimate of the structure function,
$\hat{SF}(\tau')$, at the lag $\tau'=\sum\Delta t_{i,j}/n$.

For the overall {\it EUVE} light curve (Figure~\ref{fig3}), 
we estimated the SF twice. The
first time we used $\Delta \tau=1$ day, and the second time $\Delta
\tau=1$ month. The first bin value was used to estimate the SF from lag
$\sim 1$ day up to lag $\sim 18$ days. We cannot estimate the SF at higher
lags with $\Delta \tau=1$ day since the individual{\it EUVE} light curves
of \m have a maximum length of $\sim 20 $ days. Consequently, we used the
larger bin size ($\Delta \tau=1$ month) to estimate the SF at lags $> 2$
months and up to $\sim 2.5$ years (ie the time interval between the first
and last {\it EUVE} light curve of \m). 

The combined short and long bin estimated SF is plotted in Figure~\ref{fig14}
(filled circles for $\Delta \tau=1$ day and filled squares for $\Delta
\tau=1$ month). The estimated SF rises roughly logarithmically with $\tau$
at lags $<20$ days. However, at lags $> 60$ days it is relatively flat.
This flattening cannot be due to the uneven sampling of the light curve,
but it is exactly what we would expect if there is a low frequency turn
over in the power spectrum of \m.

We examine below whether the estimated SF shown in Figure~\ref{fig14} is consistent
with the assumption that the true power spectrum of \m is similar to the
model defined by equation~(5) with $2\alpha=2.14$ and $f_{b}=4\times
10^{-6}$ Hz. 

To answer this question, we performed a numerical experiment: we
created one set of 1000 simulated light curves using the method
described in appendix C of Papadakis \& Lawrence (1995).  Each
simulated light curve has a power spectrum with slope of 2.14, a knee
frequency at $4\times 10^{-6}$ Hz, and variance as the {\it EUVE}
light curve. For each one of the simulated light curves, we estimated
the SF exactly in the same way as we did with the {\it EUVE} light
curve. Then we computed the expected mean value, $<SF_{sim}(\tau)>$,
and variance, $\sigma^2_{<SF_{sim}>}(\tau)$, of all the simulated SFs at
each $\tau$. The results are plotted in Figure~\ref{fig14} (open circles and
open squares for the mean short and long bin simulated SF).  The
``errors" on the points are equal to $\pm \sigma_{<SF_{sim}>}(\tau)$.
Note that these errors would represent the $68$\% of the process realizations 
in the case of a Gaussian distribution, but, since the actual SF are {\it not}
normally distributed, they can only give us an idea of the real $1 \sigma$ 
errors of the actual SF and cannot be used in the normal $\chi^2$ estimation.
In order to quantify the comparison between $\log[<SF_{sim}(\tau)>]$ and
$\log [\hat{SF}(\tau)]$ in a statistical way,
we estimated the sum of their squared differences, i.e.  $ \sum_{i} \{
\log[<SF_{sim}(\tau_{i})>] - \log[\hat{SF}(\tau_{i})]\}^{2}$, for all the
$\tau_{i}$ for which we have estimated the SF. We refer to this sum as
``$\chi^{2}$", although its definition is different from the traditional
definition of $\chi^{2}$. In our case, $\chi^{2}=4$.  Based on the
distribution of the $\chi^{2}$ values for the 1000 simulated light curves,
the probability that the {\it EUVE} light curve of \m is a realization
of a process with a power spectrum with a knee frequency at $4\times
10^{-6}$ Hz is $\sim 40$~\%.

Finally to check whether \m structure function is consistent with a break at 
time scales longer than the range suggested by the fit of the power spectra,
 we performed a second numerical experiment: we created a new set
of 1000 simulated light curves, with $2\alpha=2.14$, variance as the {\it
EUVE} light curve variance, and a knee frequency at $10^{-7}$ Hz. As
above, for each simulated light curve we estimated the SF, and then
computed $<SF_{sim}(\tau)>$, and $\sigma_{<SF_{sim}>}(\tau)$. The results
for the mean short and long bin simulated SF are plotted in Figure~\ref{fig14}
(dotted triangles). At large lag values the mean simulated SF agrees well
with the SF of \m. This is not surprising since, as we mentioned at the
beginning of this Section, we expect that $SF \sim 2\sigma^{2}$ in the
limit when $\tau$ is larger than $1/f_{b}$ ($10^{7}$ s in this case).
However, the agreement between the mean simulated SF and the \m SF at
short lags is not good. The $\chi^{2}$ this time is equal to 183. Based
on the distribution of the $\chi^{2}$ values for the 1000 simulated light
curves, the probability that the {\it EUVE} light curve of the \m is a
realization of a process with a power spectrum with a knee frequency at
$10^{-7}$ Hz is $<0.1$~\%.

We conclude that the SF of \m (Figure~\ref{fig14}) 
is entirely consistent with the
results from the power spectrum analysis (Section~3); the slope of the \m
power spectrum is $\sim -2.1$ and there is a knee frequency at $\sim
4\times 10^{-6}$ Hz, below which the spectrum flattens; in fact we found that it is 
highly unlikely that this knee frequency will be smaller than $10^{-7}$
Hz. 

\section{Discussion}

\subsection{Summary of results}

All the {\it EUVE} light curves of \m show significant variations
(Figures~\ref{fig4}--\ref{fig8}). One interpretation is that they 
are the result of  the superposition of flares
whose duration, roughly estimated from the same figures, is about 3
days; as an example we quote the 1994 lightcurve (Figure~\ref{fig4}) where the
source flux declines for about 3 days, rises for another $\sim 3.5$
days and stays steady to the end of the observation ($\sim 3.5$
days). We investigated the variability present in the light curves by
means of a power spectrum analysis and our results are as follows:

\begin{itemize} 

\item on average, the {\it EUVE} power spectrum of \m has a power law form
with a slope $-2.14\pm 0.28$; 

\item there is a low frequency turn over at $4\times 10^{-6}$ Hz 
(i.e. $\sim 3$ days, the 68\% confidence range is $7\times 10^{-7}$ Hz $-$ 
$8\times 10^{-6}$ Hz). We cannot
determine the shape of the power spectrum below this turn over,
nevertheless the data so far is consistent with a flattening of the power
spectrum below this frequency to a slope of zero; 

\item there is a strong indication that the power spectrum does not remain
the same, i.e. that the {\it EUVE} emission of \m is not stationary. The
DS 1996 observation shows a power spectrum with a slope $-1.1\pm 0.2$,
which is much flatter than the average slope. However, this observation is
again consistent with a flattening below the same frequency of $4\times
10^{-6}$ Hz.

\end{itemize}

Furthermore, we performed a structure function analysis of the total {\it
EUVE} light curve. The structure function of the source flattens at time
scales $> 10^{6}$ s. Based on numerical simulations, we found that the
\m SF is entirely consistent with a $2.14$ power spectrum which flattens
to a slope of zero below $4\times 10^{-6}$ Hz. 

We will discuss each of these results in more detail in the following
subsections.

\subsection{The power spectrum slope}

Our analysis shows that, on average, the {\it EUVE} power spectrum of \m
(up to $\sim 10^{-4.5}$ Hz) has a power law form with a slope of $-2.14
\pm 0.28$. No periodicities or quasi-periodicities could be found. To the
best of our knowledge, this is the first time that the power spectrum of
a radio-loud AGN  has been estimated in the EUV. 
In fact, even in the X-rays, the power spectra of radio-loud objects are
not well known. The only exception is PKS 2155-304. Its X-ray power
spectrum has been estimated in the past (Tagliaferri et al 1991, Zhang et
al 1999) to have  a slope of $\sim -2$, similar to the slope of the {\it
EUVE} power spectrum of \m.  Both objects are High frequency peak BL Lac
objects, and their {\it EUVE} and X-ray emission are thought to be due to
synchrotron radiation. 
The similarity in the power spectrum slopes could imply that 
in addition to the same emission mechanism, the same variability mechanism
operates in both objects in the two bands.\footnote{The normalized
amplitude of the PKS 2155-304 X-ray power spectrum at $10^{-5}$ Hz is 
$> 1000$ (Zhang et al. 1999), when for \m \/ the normalized amplitude 
of {\it EUVE} power spectrum at the same frequency, is $148\pm 33$ (Table~2).}

In general, red noise power spectra can result from a shot-noise model, 
i.e. from a superposition of random 
``flare"-like events,  each one with the same duration
and shape, and with an amplitude randomly distributed around a mean value.
A slope $\alpha = 2$ is expected in the case of
exponentially decaying shots, consistent with our best fit slope value.
  However, the shape of the flares in
the {\it EUVE} light curves of \m can by eye, be more complicated than that;
 they are
asymmetric and they do not have the same shape all the times. For example,
the first flare in the DS 1995 lightcurve (Figure~\ref{fig6}) 
shows a relatively
slow rise (it reaches maximum after $\sim 2$ days), and faster decline
($\sim 1$ day). Immediately after this the source rises again, reaching
now the maximum after $\sim 1$ day. Even if it is not clear what happens
after the maximum (if there is a second flare or not)  this difference in
rise times indicates that the flares are not always the same.
Probably, there are flares which rise slow and decay fast and flares with
the opposite behavior, or the complicated shapes can be the result of the 
superposition of consecutive flares.

In general, knowledge of the power spectrum slope alone cannot provide us
with much information on the variability mechanism that operates in the
source. One has to work the other way around; develop a model for the
variability mechanism, predict the shape of flares, hence the power
spectrum slope, and then compare it with the data. 

Even though more work is needed in order to
investigate if any  model can  produce light
curves with the correct power spectrum slope, amplitude and low frequency
knee,  it is interesting to notice the similarity between
the shape of some of the flares in the {\it EUVE} light curves of \m and
those predicted by Celotti et al. (1991). 
The flare at the
beginning of the DS 1995 light curve 
(Figure~\ref{fig6})  resemble by eye the
flare expected in the case of a perturbation with growing thickness in an
inhomogeneous jet (Figure~4b of Celotti et al. 1991). A different
behavior is present in the second flare in the same light curve; this
time the flux of the source continues to be large after reaching the
maximum and the amplitude of the event is smaller than in the previous
case. This flare could correspond to the case 
in Figure~4a of Celotti et al. (1991), in which they used different parameters 
for the stationary jet structure.

\subsection{The low frequency turn over}

The presence of a low frequency knee, with a value that remains constant, 
within the error bars,
in all the {\it EUVE} power spectra of \m, indicates the presence of a
``characteristic time scale" ($t_{var}$) in the system. This time scale
determines the ``memory" of the system, i.e. the variability mechanism
operates in such a way so that, whatever the origin and nature of
variations, they ``last"  typically for a period which is, on average,
equal to $t_{var} \sim 1/(4\times10^{-6})$ Hz $\sim 3$ days.  For example,
according to the simple shot noise model mentioned above, the ``memory" of
the system is determined by the duration of the flares. If, on average,
they last for $t_{var}$ days, a low frequency knee at $\sim 1/t_{var}$ is
expected. As we mentioned in the beginning of this section, visual
inspection of the \m {\it EUVE} lightcurves implies that the flares in
them last for $\sim 3$ days. 

The fact that the low frequency knee remains constant implies that
$t_{var}$ is determined by a characteristic of the source that does
not change with time. Such a permanent characteristic might be the
size of the {\it EUV} emitting region. If that is the case, then an upper
limit to the source size at rest frame, $R$, is given by
$R=t_{var} \times (\delta c)$, where $\delta$ is the Doppler factor. For
$t_{var} \sim 3$ days and a typical value of $\delta \sim 10$, the
source size should be less than $\sim 8\times 10^{16}$ cm. This value
is similar to the value of $4.8\times 10^{16}$ cm and $4.7\times
10^{16}$ cm that Chiaberge \& Ghisellini (1999) and Mastichiadis \&
Kirk (1997) find from theoretical modeling of multifrequency energy
spectra of \m.

Recently, Maraschi \etal\ (1999) proposed the e-folding times scales
of 10.7$\times10^4$ s, 7.1$\times10^4$ s, and 5.3 $\times10^4$ s at 1,
5 and 15 keV respectively for a large flare observed by {\it
BeppoSAX}. These time scales were calculated from a simple arbitrary
exponential fitting to a single event in the {\it BeppoSAX} light
curve.  If they represent indeed a characteristic time scale of the
system at these energies, then this time scale decreases with
energy. The time scale derived from EUV data is 2.6$\times10^5$, i.e.
larger than the time scales in the X-ray band.  If the time scale is
associated with the source size, then these numbers are consistent
with models in which the source size is expected to decrease with
energy. A more detailed comparison has to wait for a proper statistical
analysis of the X-ray data.

An indication of a break at $\sim 0.5$ days is present also in
the structure function analysis performed on {\it Beppo-SAX} 2000
observation of MRK~421 (Zhang Y. H. 2000 private communication).
Assuming that the bulk of {\it Beppo-SAX} emission is at $\sim 3-4$~keV,
the 6 times longer EUV ($\sim 0.1$ keV) break could be explained with the
synchrotron cooling of the same electron population emitting in the 
{\it Beppo-SAX} range.

\subsection{Evidence of non-stationarity: DS 1996 data}

As is evident from Figure~9 and Figure~12, the slope of DS 1996 power spectrum
 (open circles)
is flatter than the others, showing the non stationary nature of \m \/
emission.
The appearance of the DS 1996 light curve (Section 2.2.4) agrees well with
the power spectrum of that light curve. The flatter slope of the DS 1996
power spectrum indicates larger amplitude variations at high frequencies,
i.e. lots of short time variations, consistent with the presence of a
number of short lived flares, and the absence of any long ``trend" in the
light curve itself. Changes in the power spectrum indicates that some of
the variability mechanism characteristics change with time. 
PKS 2155-304 also shows evidence for non-stationarity in its X-ray light
curves (Zhang et al 1999). \m\/ has shown evidence of
``non-stationarity" in the past: Fossati et al 2000, in a simultaneous TeV
and X-ray observation, found that the hard X-rays were
delayed with respect to the soft X-rays, opposite to the behaviour 
previously observed  (Takahashi et al 1996). 

In the context of the shot-noise model, in DS 1996 power spectrum:
(a) the different slope indicates a difference in the flare shape;
(b) the roughly similar amplitude  (Figure~\ref{fig9}) 
suggests that probably both the number of flares per unit time and their 
amplitude has not changed drastically during DS 1996 observation;
(c) the knee at the same frequency, within the errors, of the other power 
spectra indicates that the duration of DS 1996 flares is not changed.\\
To the eye, though the flares in the DS 1996 light curve appear to be 
relatively short-lived. 
We speculate that the DS 1996 flares consist of two parts: one with a fast
decay/rise evolution, and another one with a much slower evolution, not
seen directly in the light curve, due to the blending of consecutive flares.

\section{Conclusions}

The extremely large {\it EUVE} database on \m\/ ($\sim 1100$ ks) offers a so
 far unique  opportunity to study the variability properties of a BL Lacertae
 object at high energies. 
We analyzed \m \/ light curves by means of the power spectrum and  
structure function techniques; in a power spectrum with slope $-2.14\pm 0.28$ 
we find indication, confirmed by the structure function analysis, of a break
 at $\sim 3$ days. This is the first time that a break in the power spectrum 
of a BL Lacertae object is found.
We also find  evidence of non-stationarity for \m \/ EUV emission.

The few  
existing models on BL Lacertae objects variability (Celotti et al. 1991, 
Chiaberge \& Ghisellini 1999, Georganopoulos \& Marscher 1998), 
developed for multifrequency analysis, try to explain the 
amplitude and delays observed among the various energy bands.
The single energy approach, as the one proposed in this paper, is at least as 
promising as the multifrequency approach, but no predictions on the power 
spectrum  and on the structure function shapes (the most widely used analysis 
methods to  quantify the variability of an object) 
have never been derived from any of the current models.\\
It is beyond
the scope of this paper to investigate in detail whether the model flares
that have been computed so far can reproduce the results from the analysis
of the {\it EUVE} light curves of \m and we reserve this for a future paper.
 However, such a quantitative work,
will be very interesting since it may be able to distinguish between
different models that have been proposed to explain the high frequency
emission of blazars.

\acknowledgments

We would like to thank Drs. A.~Celotti, M. Elvis, H. Marshall and L. Maraschi
 for useful discussions. 
We thank the \euve\ science team
and in particular Dr. Roger Malina for their logistical support for
our continuing analysis effort on \euve\ data.  This research has made
extensive use of the High Energy Astrophysics Science Archive Research
Center Online Service (HEASARC), provided by the NASA-Goddard Space
Flight Center, of the NASA/IPAC Extragalactic Database (NED) operated
by the Jet Propulsion Laboratory, Caltech, under contract with NASA
and of NASA's Astrophysics Data System Abstract Service.  
This work was supported by AXAF Science Center NASA contract NAS 8-39073
and by NASA grants NAG 5-3174 and NAG 5-3191.
Finally, IC thanks the little Ginevra Biaggi for the 4 months long maternity leave
during which the bulk of this paper was written.

\clearpage

\figcaption{Comparison of the effective areas of the Deep Survey
(solid line) and scanner A (dashed line) instruments on board the
\euve\ satellite.\label{fig1}}

\figcaption{Simulated count rates in the DS (solid line) 
and Scanner~A (dashed line) instruments assuming an absorbed power law
model with spectral index $\alpha=2$ The ratio of the count rate at
the peak wavelength for the two instruments was used as rescaling
factor to convert the Scanner~A count rate into a DS count
rate.\label{fig2}}

\figcaption{Combined lightcurve of all {\it EUVE} observations from 1994 to 1997 binned over one average {\it EUVE} orbit ($\sim 5544$ s).\label{fig3}}

\figcaption{April 2-12 1994 Deep Survey lightcurve binned over one average 
{\it EUVE} orbit ($\sim 5544$ s).
This light curve was also presented in Fruscione \etal\ (1996).\label{fig4}}
 
\figcaption{February 4-7 1995 Scanner A light curve (open circles) binned
over one average {\it EUVE} orbit ($\sim 5544$ s).
The filled circles represent the count rate scaled to the DS instrument (see Section~2.2.2).\label{fig5}}
 
\figcaption{Same as Figure~\ref{fig4} for the April 25 - May 13 1995 DS data. 
The data from MJD 49832 to MJD 49843 were presented in Kartje \etal\ (1997) 
and were part of a multiwavelenght campaign during wich a TeV flare 
was detected (Buckley \etal\ 1996).\label{fig6}}

\figcaption{Same as Figure~\ref{fig4} for the April 17-30 DS data (on the left) 
and the May 10-11 data (on the right). A large TeV flare (Zweerink
\etal\ 1997) occured on 7 May 1996.\label{fig7}}

\figcaption{Same as Figure~\ref{fig5} for the February 7-11 1997 Scanner A data.\label{fig8}}
 
\figcaption{Normalized power spectrum (binning factor $m=10$, see Section~3.1)
 obtained from each lightcurve binned
over 5544s (1 average {\it EUVE} orbit).
1995 DS/S observation was split into two parts to avoid the gap.
The open symbols represent DS/S observations 
(triangles for 1994, squares for the first
part of 1995, stars for the second part of 1995 and circles for 1996) 
while the filled symbols represent scanner observations 
(squares for 1995 and circles for 1997).\label{fig9}}

\figcaption{Same as Figure~\ref{fig9} for DS 1996 observation with different binnings.
Opens circles, as in Figure~\ref{fig9}, represent a binning value of $m=10$, and filled circles $m=20$ (see Section 3.1)\label{fig10}}

\figcaption{Average normalized power spectrum of \m  obtained from all 
the 5544s binned lightcurves presented in Figure~\ref{fig3} (see section 3.1). 
Different symbols represent different binnings:  $m=20$ (filled
circles) for the low frequency part  and  $m=10$ (open circles) for the 
high frequency part representing the the Poisson noise level.
The solid line shows the best power law model fit with slope
-2.14.\label{fig11}}
 
\figcaption{Power spectra of the DS 1995 (a) and DS 1996 (b)
light curves binned over 500 s (open circles). The
long dashed line in both plots indicate the expected Poisson noise
power level. Solid line shows the best fit of the model defined by
equation 5  in Section~3.2, with the slope fixed to -2.14. 
The dotted line in (a) represents the simulated power spectrum using the
 same best fit values but moving the break to $10^{-7}$~Hz; while the
 dotted line in (b) shows the best fit of the same model but with the slope
left as a free variable (see text).\label{fig12}}

\figcaption{Same as Fig~12 but for DS 1994, scanner~A 1995 and 
scanner ~A 1997 respectively.\label{fig13}}

\figcaption{Structure Function of the total {\it EUVE} light curve of \m.
 (Figure~\ref{fig3}) Filled circles correspond to the SF estimated
 with a lag bin of size 1 day and the filled squares to the SF
 estimated with a lag bin of size 1 month. Open symbols (circles and
 squares) correspond to the mean SF of 1000 simulated light curves
 with a power spectrum of slope 2.14 and a frequency turnover at
 $4\times 10^{-6}$ Hz. Dotted triangles correspond to the mean SF of
 1000 simulated light curves with a same slope power spectrum but with
 a frequency turnover at $10^{-7}$ Hz. In both cases, the errorbars
 correspond to the standard deviation of the simulated SFs around
 their mean (see Section~4 details).\label{fig14}}
\clearpage

\begin{deluxetable}{l l l l c l c c}
\tablewidth{6.9in} 
\tablecaption{EUVE Observations of Mrk~421}
\tablehead{
\colhead{Year}
& \colhead{Start}
& \colhead{End}
&
& \colhead{Instrument\tablenotemark{a}}
& \colhead{T$_{\rm exp}$\tablenotemark{b}} 
& \colhead {Count rate\tablenotemark{c}}
& \colhead {Overall Variability}\\
&&&&&\colhead{(ks)} &\colhead{($10^{-2}$ c s$^{-1}$)}
&\colhead{factor}
}
\startdata
1994 &2 Apr &12 Apr\tablenotemark{d}  &&DS   &280 &$18 \pm 0.85$  &2\\
1995 &4 Feb &7 Feb   &&ScaA &68  &$32\pm2.9$    &1.5\\
     &      &        &&     &    &$45\pm4.1$\tablenotemark{d} &\\
1995 &25 Apr &13 May &&DS   &355 &$29\pm0.96$      &3\\
1995 &25 Apr &28 Apr &(flare)\tablenotemark{e} &     &    &$37\pm 2.3$    &\\ 
1995 &29 Apr &6 May &(decay)\tablenotemark{e}  &     &    &$29\pm1.5$      &\\ 
1995 &7 May  &13 May &(quiescence) &     &    &$22\pm 1.0$    &\\   
1996 &17 Apr &30 Apr &&DS   &299 &$30\pm1.08$      &1.4\\
1996 &10 May &11 May &&DS   &3.6 &$40\pm 1.1$    &\\
1996 &7 Feb  &11 Feb &&ScaA &108 &$19 \pm 1.6$    &2\\
     &      &        &&     &    &$27\pm2.2$ &\\
\enddata
\tablenotetext{a} {``DS'' indicates that the source was observed
in the Deep
Survey photometer. ``ScaA'' indicates that the source was observed
in one of the three {\it EUVE} photometers (scanner A) during a pointing
within the Right Angle Program.}
\tablenotetext{b}{Total exposure time calculated eliminating (i) all SAA
passages (ii) all satellite daytime data (iii) all times during which
the detector was turned off (iv) all times affected by possible earth
blockage. It does include corrections for telescope vignetting,
deadtime and limited telemetry allocation (primbshing).}
\tablenotetext{c}{Average count rate in the Lexan/B filter ($\approx 60-180$ \AA). 
For the scanner observations, the second line gives the count rate 
normalized to the one in the DS assuming an average spectral
shape, to take into account the 
difference in the effective area between the two instruments.}
\tablenotetext{d}{This observation was presented in Fruscione et al. (1996)}
\tablenotetext{e}{This observation was presented in Kartje
et al.( 1997)}
\end{deluxetable}

\clearpage

\begin{deluxetable}{l l c c c c }
\tablecaption{Power spectra ``power law'' model fits for  the
DS orbit-binned light curves of MRK~421}
\tablehead{
\colhead{Year} & \colhead{Instrument} 
& \colhead {$\alpha$}
& \colhead {A}
& \colhead {No. of points}
& \colhead {$\chi^2$}
}
\startdata
1994	 	&DS&$4.02 \pm 1.42$	&$13 \pm 24$\tablenotemark{a} &3	&0.38\\
1995 (part 1)	&DS&$4.14 \pm 0.86$	&$22 \pm 15$\tablenotemark{a} &4	&0.35\\
1995 (part 2)  &DS&$2.64 \pm 1.10$	&$15 \pm 8$\tablenotemark{a} &3	&0.17\\
1996	 	&DS&$1.45 \pm 0.42$	&$13 \pm 6$\tablenotemark{a} &5	&3.85\\
All	&DS+scaA &$2.14 \pm 0.28$	&148 $\pm 33$\tablenotemark{b} &10 &5.6\\
\enddata
\tablenotetext{a} {in units of (counts/sec)$^{2}$/Hz}
\tablenotetext{b} {in units of Hz$^{-1}$}
\end{deluxetable}

\clearpage

\begin{deluxetable}{l l c c c c }
\tablecaption{Power spectra ``power Law plus low frequency turn over''
model fit for the DS and scanner A, 500 sec binned light curves
of MRK~421}
\tablehead{
\colhead{Year} 
& \colhead{Instrument} 
& \colhead {Slope$2\alpha$}
& \colhead {$f_{b}$}
& \colhead {DoF}
& \colhead {$\chi^2$}\\
&&
& \colhead {(Hz)}
&&\\
}
\startdata
1994  &DS &2.14 (fixed) & $8.4^{+27}_{-7.7} \times 10^{-6}$ &43 & 
67.8 \\
1995  &ScaA &2.14 (fixed) & $5.0^{+18}_{-4.3} \times 10^{-5}$ &14 &
28.7\\
1995  &DS &2.14 (fixed) & $2.6^{+5.6}_{-2.0} \times 10^{-6}$ &77 &
116.9 \\
1996  &DS &2.14 (fixed) & $7.7^{+4.3}_{-2.6} \times 10^{-5}$ &53 &
140.4 \\
      &   &$(1.1\pm 0.1)$ &$(1.6^{+10.}_{-1.5}\times 10^{-6}$) &(52) &
(65.5) \\
1997  &ScaA &2.14 (fixed) & $7.7^{+17}_{-5.3}  \times 10^{-5}$ &17 &
16.7 \\
\enddata
\end{deluxetable}
\clearpage


\end{document}